%% file: main.tex
\documentclass[conference]{IEEEtran}
\usepackage{graphicx}
\usepackage{comment}
\usepackage{balance}
\usepackage{xcolor}
\definecolor{linkc}{rgb}{0.1,0.1,.8}
\definecolor{darkgreen}{rgb}{0,0.5,0}
\definecolor{midblue}{rgb}{0,0,0.7}
\usepackage{url}
\usepackage[bookmarks=false,colorlinks=true,citecolor=darkgreen,linkcolor=linkc,filecolor=linkc,urlcolor=linkc]{hyperref}

\usepackage{amsmath}\interdisplaylinepenalty=2500
\usepackage{algorithmic}
\usepackage[caption=false,font=footnotesize]{subfig}
\usepackage{maple} 
\usepackage{listings}
\input{listings-maple-definition.sty}
\definecolor{lstbg}{cmyk}{0.07, 0.04, 0.18, 0}
\usepackage{cleveref}
\usepackage{cite}      
\usepackage{array}     
\newcommand{\texttit}[1]{\texttt{\textit{#1}}}
\begin{document}

\author{
\IEEEauthorblockN{Stephen M. Watt}
\IEEEauthorblockA{Ontario Research Centre for Computer Algebra \\ and Cheriton School of Computer Science \\
University of Waterloo, Canada\\
\texttt{smwatt@uwaterloo.ca}}
\and
\IEEEauthorblockN{D. J. Jeffrey}
\IEEEauthorblockA{Ontario Research Centre for Computer Algebra \\ and Department of Mathematics \\
University of Western Ontario, Canada\\
\texttt{djeffrey@uwo.ca}}
}

\title{An Abstraction-Preserving Block Matrix Implementation in Maple}
\date{}
\maketitle
\begin{abstract}
A Maple implementation of partitioned matrices is described.  
A recursive block data structure is used, with all operations preserving the block abstraction.
These include constructor functions, ring operations such as addition and product, and inversion.
The package is demonstrated by calculating
the $PLU$ factorization of a block matrix.
\end{abstract}
\maketitle
\section{Introduction}
\label{sec:introduction}
Partitioning a matrix into blocks is an elementary concept in linear
algebra \cite{OlverApplLinAlg2018, Anton11}.
It is supported in all major  mathematical software systems,
but typically only as a means for building a matrix or specifying
submatrices -- a matrix is flattened before operations can be performed
with it.
In more advanced studies, however, the abstraction of a block matrix is
often used to present and analyse algorithms \cite{Trefethen}.
Therefore, an ability to work with block matrices as computational
objects is desirable. In addition to theoretical studies, block matrices
offer potential benefits for computation:
\begin{itemize}
    \item dense, sparse and structured matrices can be represented with reasonable space efficiency~\cite{gargantini-quadtree,abdali-wise-quadtrees},
    \item block matrices provide a middle ground that avoids pathological communication bottlenecks in row-major or column-major code~\cite{dongara-scalability}, and
    \item multiplication and related algorithms can have improved computational time complexity~\cite{strassen-mult}.
\end{itemize}
The basic algebra of block matrices is covered in standard references \cite{Reams}, and a theoretical basis for implementing operations has been outlined in \cite{watt-recursive-block-algos}. 

An implementation is described here that preserves and respects the block structure for all the usual matrix operations, including inversion.
The functions are demonstrated by using them to compute a matrix $PLU$ factorization.  

The implementation is in Maple, allowing the ultimate matrix elements to be symbolic expressions.
To work efficiently with block matrices with floating point or finite field entries, a compiled language such as \texttt{C++}, Rust, Julia or Aldor~\cite{watt-aldor} may be used.  All of these offer parametric types and operator overloading, allowing elegant implementations.

The remainder of the paper is organized as follows:
Section~\ref{sec:datastructure} describes the Maple data structure we use for recursive block matrices and some of the considerations in the choice.
Section~\ref{sec:module} presents a Maple module used to encapsulate the data representation and describes the operations provided.
Section~\ref{sec:plu} shows how the block operations may be used to compute $PLU$ decomposition of square non-singular matrices.
Section~\ref{sec:example} gives examples of using the package and
Section~\ref{sec:future} outlines some directions for future work.
Finally, \ref{sec:conclusions} gives some brief conclusions.

\section{Data Structure}
\label{sec:datastructure}
We describe here the data structure we have used for block matrices.
We have used an expression tree structure with the interior nodes being symbolic function applications, \texttt{\_BM(\textit{kind}, \textit{er}, \textit{ec}, \textit{val})}.  This is represented as a pair with the first element being a pointer to the name \texttt{\_BM} and second element being a pointer to a four element ``expression sequence'' which is basically four consecutive words in memory plus a header.

The first element, \texttt{\textit{kind}}, is a name being one of
\texttt{zero}, 
\texttt{scalar}, 
\texttt{matrix} or 
\texttt{rblock}.
The meaning of these is described below.

The second and third elements, \texttit{er} and \texttit{ec}, are integers giving the number of leaf rows and columns, respectively.  For example, the block matrix
\[
\left [ \begin{matrix}
   \left [ \begin{matrix} 11 & 12 \\ 13 & 14 \end{matrix} \right ] &
   \left [ \begin{matrix} 15 & 16  & 17\\ 18 & 19 & 20 \end{matrix} \right ]
   \\
   \\
   \left [ \begin{matrix} 21 & 22 \end{matrix} \right ] &
   \left [ \begin{matrix} 23 & 24 &25 \end{matrix} \right ]
\end{matrix} \right ]
\]
would have 
\texttt{\textit{er}} $= 3$ and 
\texttt{\textit{ec}} $= 5$.

The meaning of the fourth element, \texttit{val}, depends on the value of \texttit{kind}:
\begin{itemize}
    \item When \texttit{\textit{kind} = zero}, the block is interpreted as a zero matrix whose entries are not explicitly stored.
    The value of \texttt{\textit{val}} is always the integer $0$.
    \item When \texttt{\textit{kind} = scalar}, the block is interpreted
    as a  diagonal matrix equal to 
    $\text{\texttit{val}}$ times the 
    $\text{\texttit{er}}\times\texttit{er}$ 
    identity.  
    To be well-formed, it is required that $\text{\texttit{er}} = \text{\texttit{ec}}$.
    \item  When \texttt{\textit{kind} = matrix}, \texttit{val} is
    a $\text{\texttit{er}}\times\texttit{er}$ Maple object of type \texttt{Matrix} whose entries are leaf elements of the block matrix.
    \item  When \texttt{\textit{kind} = rblock}, \texttit{val} is
    a Maple object of type \texttt{Matrix} whose entries are themselves block matrices. (The name connotes ``recursive block''.)
    
    To be well formed, two conditions are required:
    \begin{itemize}
    \item Each of the blocks in a given row must have the same \texttit{er}, and the sum of their \texttit{ec} values must equal the \texttit{ec} value of the whole block matrix. 
    \item Each of the blocks in a given column must have the same \texttit{ec}, and the sum of their \texttit{er} values must equal the \texttit{er} value of the whole block matrix. 
    \end{itemize}
\end{itemize}
There are several optimizations that could be done to make this representation more space efficient, but the above representation is convenient for development purposes.  For example, it would be possible to use the symbolic \texttit{kind} names in place of \texttt{\_BM} and have different length expression sequences for each kind, or it would be possible to have a \texttit{kind} for general diagonal matrices, and so on.
It would also be possible to omit the \texttit{er} and \texttt{ec} fields
in some of the kinds, since these can sometimes be determined by inspection, but having them always present gives more uniform code and avoids computation.

\section{Module}
\label{sec:module}
All of the operations on block matrices are collected in a Maple module that hides the representation and presents an abstract interface.
The following operations are provided.

\subsection{Construction operations:}
\begin{itemize}
    \item \texttt{BM(\textit{vals})}
    constructs a block matrix from a list of lists or an object of type \texttt{Matrix}.  
    The \texttit{kind} is \texttt{rblock} if all the entries of \texttit{vals} are block matrices.  
    Otherwise the \texttit{kind} is \texttt{matrix}.
    
    Finer control is provided by the constructors below.
    
    \item \texttt{zeroBM(\textit{er}, \textit{ec})}
    gives \texttit{kind} \texttt{zero} of dimension
    $\text{\texttit{er}}\times\text{\texttit{ec}}$.
    
    \item \texttt{scalarBM(\textit{er}, \textit{s})}
    gives \texttit{kind} \texttt{scalar} of dimension
    $\text{\texttit{er}}\times\text{\texttit{er}}$,
    interpreted as a matrix with diagonal entries equal to \texttit{s} and off-diagonal entries equal to 0.
    
    \item \texttt{matrixBM(\textit{m})}
    gives \texttit{kind} \texttt{matrix} with entries given by the matrix \texttit{m}.
    
    \item \texttt{rblockBM(\textit{m})}
    gives \texttit{kind} \texttt{rblock} with entries themselves being block matrices given by the elements of matrix \texttit{m}.

    \item A number of other specialzied constructors.
\end{itemize}
\subsection{Operations to abstract the type:}
\begin{itemize}
\item \texttt{`type/BM`(\textit{B})} tests whether \texttit{B} is a block matrix,  returning \texttt{true} or \texttt{false}.  This allows Maple statements of the form \verb+if type(A, 'BM') then ..+ so block matrices participate in Maple's structured type system.
\item
    \texttt{nEltRows(\textit{B})} and
    \texttt{nEltCols(\textit{B})} give the number of leaf element rows and columns.
\item
    \texttt{elt(\textit{B},\textit{r},\textit{c})}
    return the $(\text{\texttit{r}}, \text{\texttit{c}})$ leaf element, \textit{i.e.} 
    when \texttit{B} is viewed as a un-partitioned matrix.
\item
    \texttt{nBlockRows(\textit{B})} and
    \texttt{nBlockCols(\textit{B})} gives the number of rows of blocks and columns of blocks.  
\item
    \texttt{block(\textit{B},\textit{r},\textit{c})}
    return the $(\text{\texttit{r}}, \text{\texttit{c}})$ block,
    \textit{i.e.} when \texttit{B} is viewed as a block matrix.
\end{itemize}
\noindent Operations to traverse the data structure:
\begin{itemize}
\item
    \texttt{map(\textit{f}, \textit{A})} returns a block matrix whose elements are those of \textit{A} with the function $f$ applied.
    
    Logically, if $R = \text{\texttt{map}}(f, A)$,
    then $R_{ij} = f(A_{ij})$, for 
    $1 \le i \le \text{\texttit{er}},
    1 \le j \le \text{\texttit{ec}}.$
    The resulting matrix always has the same block structure as \texttit{A}, but the kinds of the blocks can change.
    For example if $f(0) \ne 0$ then a \texttt{zero} block will have a \texttt{matrix} block as its image.
\item
    \texttt{zip(\textit{f}, \textit{A}, \textit{B})} returns a block matrix whose elements are the values of \texttit{f} applied to pairs of corresponding elements from \texttit{A} and \texttit{B}. 
    Logically, if $R = \text{\texttt{zip}}(f, A, B)$,
    then $R_{ij} = f(A_{ij}, B_{ij})$, for 
    $1 \le i \le \text{\texttit{er}},
    1 \le j \le \text{\texttit{ec}}.$
    \texttit{A} and \texttit{B} must have the same dimension and block structure and this will be the block structure of the result.
    As with \texttt{map}, the kinds of the blocks can change.
    The implementation handles combinations of different kinds of blocks. 
    The following combinations of \texttt{kind}s are allowed:
    \begin{itemize}
    \item \texttt{zero} with any kind and any kind with \texttt{zero}
    \item \texttt{scalar} with any kind and any kind with \texttt{scalar}
    \item \texttt{matrix} with \texttt{matrix} and
    \item \texttt{rblock} with \texttt{rblock}.
    \end{itemize}
\item
    \texttt{blockKind(\textit{B})} and \texttt{blockVal(\textit{B})}  return the \texttit{kind} and \texttit{val} fields \texttit{B}
    respectively.
    These are lower-level operations, not intended for external use, but which are required when traversing or combining block matrix data structures such as with \texttt{map} or \texttt{zip}.
\end{itemize}
\subsection{Matrix ring operations:}
\begin{itemize}
\item
    \texttt{plus(\textit{A}, \textit{B})}  and
    \texttt{`minus`(\textit{A}, \textit{B})} 
    compute 
    $\text{\texttit{A}} + \text{\texttit{B}}$ and
    $\text{\texttit{A}} - \text{\texttit{B}}$ respectively.
    The same shape rules apply as for \texttt{zip}.
    The grave characters ``\verb+`+'' are required because \texttt{minus} is a keyword in Maple meaning set difference.
\item
    \texttt{neg(\textit{\textit{A}})} computes
    $\text{-\text{\texttit{A}}}$.  
    The shape rules are as for \texttt{map}.
\item
    \texttt{times(\textit{A}, \textit{B})} computes $A \cdot B$.
    The inputs must be structured such that the required products of blocks
    is defined.  
    The same combinations of blocks are allowed as for \texttt{zip}.
    At present, only classical matrix multiplication is used, though
    it would be straightforward to use Strassen recursive multiplication beyond a given size.
\item
    \texttt{hermTrans(\textit{A})} computes the Hermitian transpose of \texttit{A}, that is, logically, 
    $\text{\texttit{A}}_{ij} =  \overline{\text{\texttit{A}}_{ji}}$,
    where $\overline z$ denotes complex conjugation.
    The blocks of the result are of the same kinds as transposed blocks of \texttit{A}.
\end{itemize}
\subsection{Inversion-related operations:}
\label{sec:inv-related}
\begin{itemize}
\item
    \texttt{inv(\textit{M})}
    computes the multiplicative inverse of \texttit M, that is $\texttit{A}^{-1}$.  
    If the block matrix is singular, then \texttt{FAIL} is returned.
    At present only $1 \times 1$ and $2 \times 2$ blocks are handled,
    though it would be straightforward to handle more rows and columns of blocks by grouping them.\\[0.75ex]
    First, \texttt{tryInv} is called, attempting to compute the inverse assuming that the $(1,1)$ block is non-singular.  
    If that fails, the more general method~\cite{watt-block-inv} provided by \texttt{invByMTM} is used.
    \\[0.75ex]
    In principle, if the $(1,1)$ block is singular, it would be possible
    to permute the blocks and try again to see if any other block is non-singular.  
    But whether the multiple tries would give a better expected execution time less than the general method may depend on the element type.
    An alternative would be to precondition by multiplying with a random
    invertible matrix.
    
    \item
        \texttt{tryInv(\textit{M})}
        attempts to invert \texttit{M} using
    \[
    \left [ \begin{matrix} A & B \\ C & D \end{matrix} \right ] ^{-1} =
    \left [ \begin{matrix}
            A^{-1}+A^{-1}BS_A^{-1}CA^{-1} &
            -A^{-1}BS_A^{-1} \\
            -S_A^{-1}CA^{-1} &
            S_A^{-1}
            \end{matrix} \right ]
    \]
    where $S_A = D - C A^{-1} B$ is the Schur complement of $A$.
    This may fail by having $A$ singular, even if the whole matrix is invertible.
    However, if the whole matrix and $A$ are both invertible, then $S_A$ will be as well.
\item
    \texttt{invByMTM(\textit{M})} computes the inverse of \texttit M using
    \[
    M^{-1} = (M^\dagger M)^{-1} M^\dagger,
    \]
    where $M^\dagger$ is the Hermitian transpose of $M$.  For formally real or complex element rings, $M^\dagger M$ will have invertible principal minors. In particular its $(1,1)$ will be invertible by \texttt{tryInv}.
\item
    \texttt{schurComp(\textit{M}, \textit{i}, \textit{j}, \textit{inv},...)}
    computes the Schur complement of the $(i,j)$ block in \texttit{M}.
    The inversion required to compute the Schur complement is performed using the \texttit{inv} functional parameter.
    Optional extra arguments allow the return of some of the products used in computing the Schur complement. 
\item
    \texttt{PLU\_Decomp(\textit{M})} computes the $PLU$ decomposition of $M$, if possible, or returns $FAIL$. 
    The result is a triple $(P, L, U)$ of block matrices, where
    $P$ represents a permutation matrix,
    $L$ represents a lower triangular with diagonal elements equal to 1, and
    $U$ represents an upper triangular matrix.
    The block matrices $L$ and $U$ are triangular as matrices of elements, not only as matrices of blocks.  For example, the $(1,1)$ block of $L$ is also lower triangular, and recursively.
\end{itemize}

\section{$PLU$ Decomposition}
\label{sec:plu}
One of the goals of this work was to implement the ideas of~\cite{watt-recursive-block-algos}, and, in particular, the description of $PLU$ decomposition of non-singular block matrices using only block operations.
As described in Section~\ref{sec:inv-related}, the $PLU$ decomposition of a matrix $M$ is a triple, $(P, L, U)$ such that $M = P L U$, $P$ is a permutation matrix, $L$ is lower triangular and $U$ is upper triangular.
We require $L$ and $U$ to be triangular element-wise, not just block-wise, and we take the disambiguating convention that the diagonal of $L$ contains 1s.

For the present, we take the simplifying assumption that $M$ is a square block matrix and the leading principal minors of $M$ are non-singular.
Then, as shown in~\cite{watt-recursive-block-algos}, if
\[
M = \left [ \begin{matrix} A & B \\ C & D \end{matrix} \right ],
\]
and we seek $M = L U$ such that
\begin{align*}
    L &= \left [ \begin{matrix} L_1 & 0 \\ X & L_2 \end{matrix} \right ] &
    U &= \left [ \begin{matrix} U_1 & Y \\ 0 & U_2 \end{matrix} \right ]
\end{align*}
with $L_1$ and $L_2$ lower triangular and $U_1$ and $U_2$ upper triagular,
we may compute
\begin{align*}
    L_1 U_1 &= A          &&\text{recursively} \\
    X &= C U_1^{-1} \\
    Y &= L_1^{-1} B \\
    L_2 U_2 &= D - X Y    &&\text{recursively}. \\
\end{align*}
\input{fig_plu}%
The implementation of the core $2\times 2$ \texttt{rblock} case is shown in Figure~\ref{fig:plu}.
Notice that we are able to use the functions described so far to provide
relatively easy to read, natural code.

\section{Examples}
\label{sec:example}
We now give examples of using the \texttt{BlockMatrix} module.
In order to have results that can be shown in the article, we use small examples with simple elements.
The first example, shown in Figure~\ref{fig:eg_inv}, inverts a block matrix all of whose blocks are singular.  
\input{fig_eg_inv}
The second example, shown in Figure~\ref{fig:eg_plu}, performs a $PLU$ factorization.

\section{Future Work}
\label{sec:future}
There are a number of additional operations and tidying up that would be required for a generally useful block matrix package, including
\begin{itemize}
    \item re-organizing block matrices to desired compatible shapes
    \item inversion of other than $1\times1$ and $2\times 2$ block matrices
    \item $PLU$ decomposition of non-square matrices and matrices with singular principal minors
    \item matrix norms
    \item eigenvalue computation
    \item and many others.
\end{itemize}
It would also be useful to examine the trade-offs in computing inverses --- when to try to pivot blocks to give a non-singular $(1,1)$ block
 \textit{versus} going to the more general $(M^\dagger M)^{-1} M^\dagger $ method
 \textit{versus} preconditioning by a matrix other than $M^\dagger $.

 Finally, it would be useful to experiment with a compiled programming language with parametric polymorphism and overloading to see how competitive these methods can be on larger numerical examples.

\section{Conclusions}
\label{sec:conclusions}
We have described a Maple implementation of recursive block matrices and operations on them and we have shown that many of the operations of linear algebra can be performed using block operations only, without breaking the block abstraction.  The whole package is about 535 lines of Maple code.

\input{fig_eg_plu}
\IfFileExists{IfExistsUseBBL.bbl}{%

\input{main.bbl}
}{%
\bibliographystyle{plain}
\bibliography{main.bib}
}
\end{document}

%% file: fig_plu.tex
\begin{figure}[t]
\begin{lstlisting}
a := block(M,1,1); b := block(M,1,2);
c := block(M,2,1); d := block(M,2,2);

ra, ca := nEltRows(a), nEltCols(a);
rb, cb := nEltRows(b), nEltCols(b);
rc, cc := nEltRows(c), nEltCols(c);
rd, cd := nEltRows(d), nEltCols(d);

# Check shape.
orElse(ra = ca and rd = cd,
       "Principal block not square");
orElse(ra = rb and rc = rd and
       ca = cc and cb = cd,  "Bad shape");
n1 := ra;
n2 := rd;

# Compute decomposition recursively.
p1, l1, u1 := PLU_Decomp(a);           # n1 x n1 all
l1inv      := inv(l1);                 # n1 x n1
u1inv      := inv(u1);                 # n1 x n1
x          := times(c, u1inv);         # n2 x n1
y          := times(l1inv, b);         # n1 x n2
y          := times(inv(p1),y);        # n1 x n2
t          := `minus`(d, times(x,y));  # n2 x n2
p2, l2, u2 := PLU_Decomp(t);           # n2 x n2 all

# Zero blocks of the neede sizes.
z12 := zeroBM(n1,n2); z21 := zeroBM(n2,n1);

# Return P, L, U
BM([[p1, z12],[z21,p2]]),  # P
BM([[l1,z12],[x,l2]]),     # L
BM([[u1,y],[z21,u2]])      # U
\end{lstlisting}
\caption{$2\times2$ \texttt{rblock} case of \texttt{PLU\_Decomp}}
\label{fig:plu}
\end{figure}%

%% file: fig_eg_inv.tex
\begin{figure}[t]
\centering
\includegraphics[width=.48\textwidth]{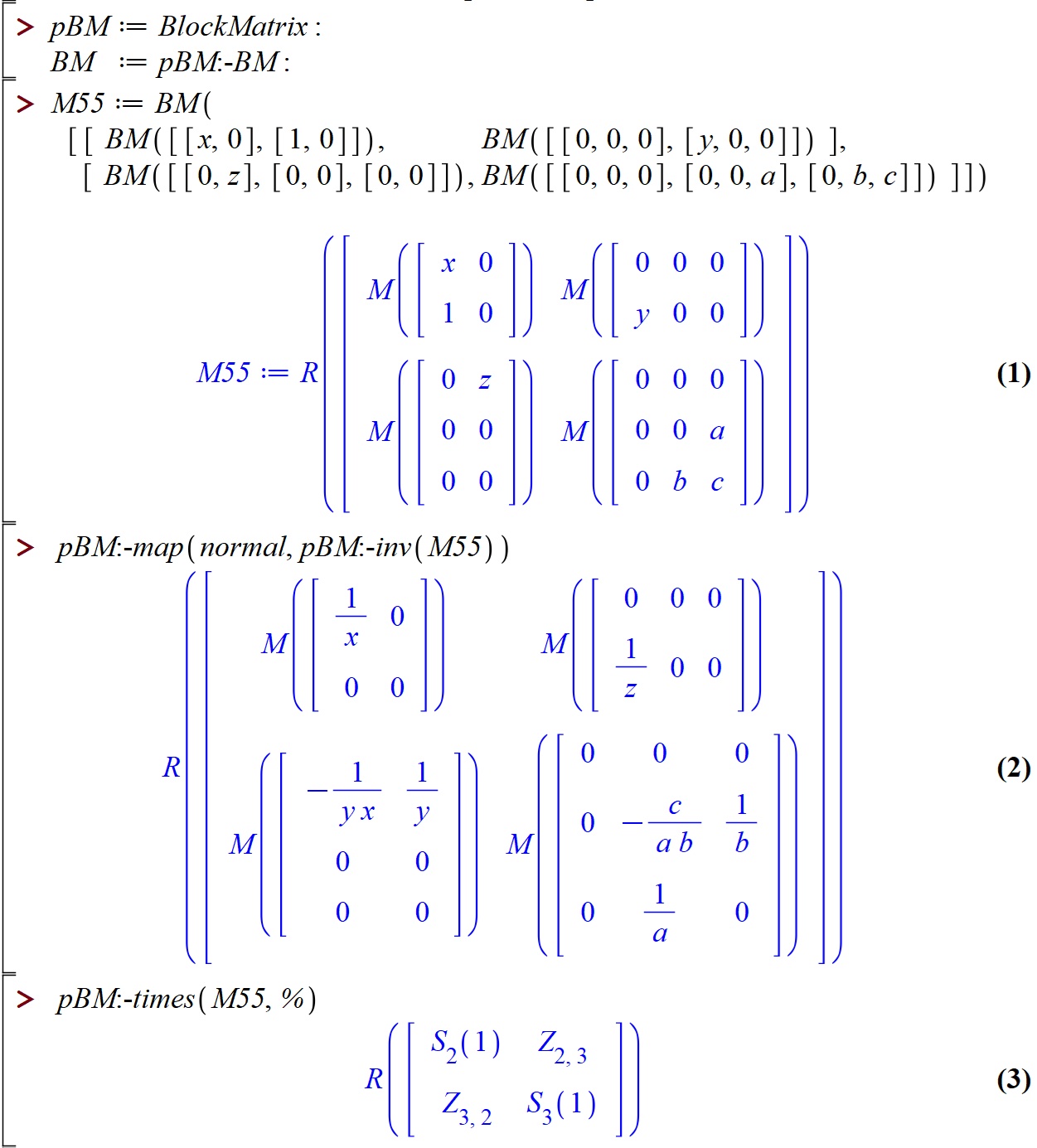}
\caption{Inversion of a block matrix whose blocks are all singular.  The symbols $R,M,Z,S$ respectively indicate recursive, matrix, zero and scalar nodes. The last line verifies the matrix and inverse multiply to give an identity matrix.}
\label{fig:eg_inv}
\end{figure}%

%% file: fig_eg_plu.tex
\begin{figure}[t]
\centering
\includegraphics[width=.48\textwidth]{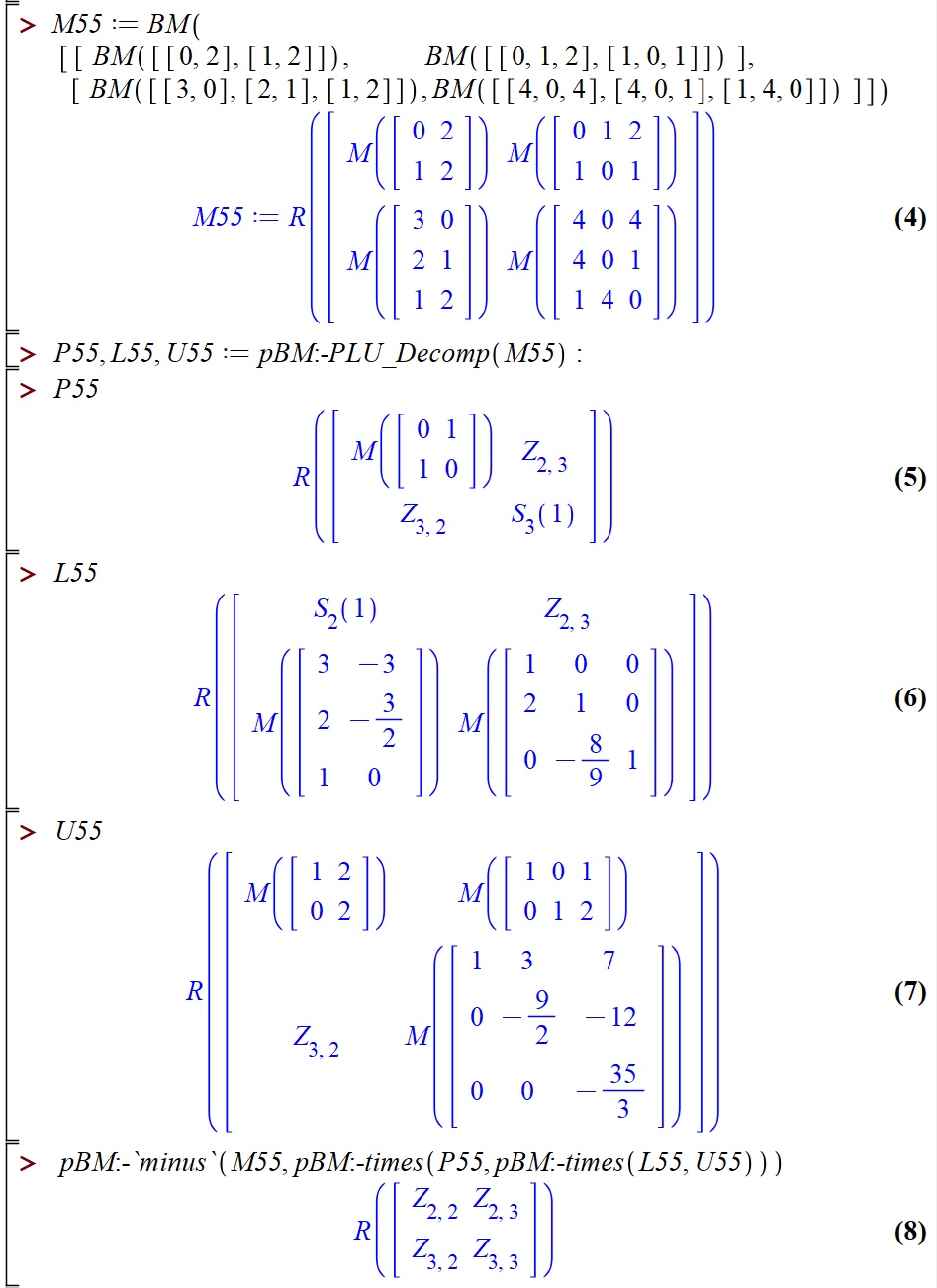}
\caption{$PLU$ decomposition of a block matrix and check of result}
\label{fig:eg_plu}
\end{figure}%